\pdfoutput=1                            
\def\lcversion{}					    

\documentclass[nonote,noversion,squeezed]{cdcarticle}

\usepackage[utf8]{inputenc}				
\usepackage[english]{babel}				
\usepackage[hidelinks]{hyperref}		

\usepackage{enumitem}					
\usepackage{cite}						

\usepackage[margin=3mm,font=small,labelfont=bf]{caption}

\usepackage{graphicx}
\graphicspath{{./graphics/}}

\usepackage{math}						
\usepackage{lipsum}						

\usepackage{arydshln}   

\newcommand{\bsmat}[1]{\left[ \begin{smallmatrix} #1 \end{smallmatrix} \right]}

\DeclareMathOperator{\sym}{sym}

\newcommand{\squeeze}[1]{\addtolength{\arraycolsep}{-#1}}
\begin{document}

\title{Direct Synthesis of Iterative Algorithms With Bounds on Achievable Worst-Case Convergence Rate}

\author{Laurent Lessard$^1$ \and  Peter Seiler$^2$}
\note{}

\maketitle

\footnotetext[1]{L.~Lessard is with the Department of Electrical and Computer Engineering and the Wisconsin Institute for Discovery, University of Wisconsin--Madison, USA.
\texttt{laurent.lessard@wisc.edu}}
\footnotetext[2]{P.~Seiler is with the Department of Electrical Engineering and Computer Science, University of Michigan, USA.
\texttt{pseiler@umich.edu}}


\begin{abstract}
Iterative first-order methods such as gradient descent and its variants are widely used for solving optimization and machine learning problems. There has been recent interest in analytic or numerically efficient methods for computing worst-case performance bounds for such algorithms, for example over the class of strongly convex loss functions. A popular approach is to assume the algorithm has a fixed size (fixed dimension, or memory) and that its structure is parameterized by one or two hyperparameters, for example a learning rate and a momentum parameter. Then, a Lyapunov function is sought to certify robust stability and subsequent optimization can be performed to find optimal hyperparameter tunings.
In the present work, we instead fix the constraints that characterize the loss function and apply techniques from robust control synthesis to directly search over algorithms. This approach yields stronger results than those previously available, since the bounds produced hold over algorithms with an arbitrary, but finite, amount of memory rather than just holding for algorithms with a prescribed structure.
\end{abstract}


\section{Introduction}\label{sec:intro}

First-order methods are a widely used class of iterative algorithms for solving optimization problems of the form
$\min_{x\in\R^n} f(x)$,
where $f$ is continuously differentiable and we only have access to first-order (gradient) measurements of $f$.
An example of a first-order method is the \textit{heavy ball} method~\cite{polyak1987introduction}, which uses iterations of the form
\begin{equation}\label{eq:heavyball}
x_{k+1} = x_k - \alpha \grad f(x_k) + \beta (x_k - x_{k-1}).
\end{equation}
Here, $\alpha$ and $\beta$ are parameters to be tuned. When $\beta=0$, we recover the well-known \textit{gradient descent} algorithm. Algorithms such as~\eqref{eq:heavyball} are called \textit{accelerated} because when they are appropriately tuned, they can achieve faster convergence than by using gradient descent alone.

To quantify convergence, we typically make assumptions about $f$, and then bound the worst-case convergence rate of the algorithm over all functions that satisfy the assumptions. For example, suppose $f$ is \textit{strongly convex} and has \textit{Lipschitz-continuous gradients}, which means there exists $0 < m \le L$ such that \cite[Thm.~2.1.12]{NesterovBook}
\begin{multline}\label{eq:convexity_assumptions}
\bmat{\grad f(x)\!-\!\grad f(y) \\ x-y}^{\!\tp}\!\!
\squeeze{3pt}\bmat{ -2mL & L+m \\ L+m & -2 }\!\!
\bmat{\grad f(x)\!-\!\grad f(y) \\ x-y} \ge 0 \\
\text{for all }x,y\in\R^n.
\end{multline}
If $x^\star \defeq \argmin_{x\in\R^n} f(x)$, then gradient descent with $0 \le \alpha \le \tfrac{2}{L+m}$ satisfies~\cite[\S2.1.5]{NesterovBook}
\begin{align}\label{eq:gradrate}
\norm{x_{k+1}-x^\star} \le
\sqrt{ 1 - \tfrac{2\alpha m L}{L+m}}
\, \norm{x_k-x^\star}.
\end{align}
In other words, the sequence of iterates $\{x_0,x_1,\dots\}$ converges to the global optimizer $x^\star$ at a geometric rate. The fastest rate bound is $\tfrac{L-m}{L+m}$, achieved when $\alpha=\tfrac{2}{L+m}$.

In an attempt to generalize these types of results and better understand the behavior of iterative algorithms and how they should be tuned, recent works have focused on a dynamical systems interpretation of optimization algorithms. In other words, the iterate sequence $\{x_k\}$ corresponds to the state, and the index $k$ serves the role of \textit{time}. Viewed in this light, convergence of an algorithm is equivalent to the stability of the associated equilibrium point of the dynamical system.


A recently proposed approach uses robust control to analyze worst-case convergence rates~\cite{ralg_ACC,lessard16,diss_ICML,cint_ACC}. Here, the algorithm itself plays the role of the controller, and the gradient mapping $\grad f$ is the uncertain plant. This viewpoint has also been extended to analyze operator-splitting methods~\cite{iqcadmm_ICML} and distributed optimization algorithms~\cite{distrop_Allerton}. 

\paragraph{Main contributions.}
In the aforementioned robust control approach, the results obtained are general in the sense that they hold across all choices of parameters as in~\eqref{eq:gradrate}, but they are fundamentally constrained by the structure of the algorithm; the particular parametric form of the iterates such as~\eqref{eq:heavyball}.

In the present work, we apply tools from robust control synthesis to derive performance bounds that are \textit{independent} of parametric forms such as~\eqref{eq:heavyball} and hold over broad classes of algorithms. These results adapt the approach in \cite{gahinet1994linear} for the purpose of synthesizing algorithms that optimize the convergence rate.

Specifically, we consider the class of functions with
\textit{sector-bounded gradients} (a weaker version
of~\eqref{eq:convexity_assumptions}). We show that, for this function
class, gradient descent with stepsize $\tfrac{2}{L+m}$ achieves the
fastest possible worst-case convergence rate not only among all
tunings of gradient descent, but among any algorithm representable as
a linear time-invariant (LTI) system with finite state. This includes
algorithms of the form~\eqref{eq:heavyball} but where $x_{k+1}$
depends linearly on $\{x_k,x_{k-1},\dots,x_{k-\ell}\}$ for some fixed
$\ell$.

Our analysis also extends to the case where $\grad f$ is characterized by a given and fixed integral quadratic constraint (IQC). In this case, we prove that when using the so-called \textit{off-by-one IQC} introduced in~\cite{lessard16}, which is satisfied by the class of functions~\eqref{eq:convexity_assumptions}, the optimal rate over any LTI algorithm is achieved by a particular accelerated method known as the \textit{triple momentum method}~\cite{van2018fastest}. 

The remainder of the paper is organized as follows. In Section~\ref{sec:prelim}, we introduce the machinery from robust control synthesis that will be used throughout the rest of the paper. In Sections~\ref{sec:sector} and~\ref{sec:sloperestricted}, we specialize the analysis to the sector-bounded and IQC cases, deriving the optimal LTI algorithms in both instances. We conclude in Section~\ref{sec:conclusion} with some closing remarks.


\section{Preliminaries}\label{sec:prelim}

\paragraph{Notation.} We use $\succ$ and $\succeq$ to denote matrix inequality in the definite and semi-definite sense, respectively. $X^\tp$ denotes the matrix transpose. We use the short-hand notation: $\sym X \defeq X+X^\tp$ with the convention that products are expanded as $\sym XY = XY+Y^\tp X^\tp$.

\paragraph{Linear matrix inequalities.} Most of the algebraic manipulations in the sequel involve linear matrix inequalities (LMI), which have now become a widespread tool in optimization and control. We refer the reader to~\cite{boydLMI} and references therein. We will make use of three fundamental results involving LMIs. The first is a basic property of the Schur complement, which can convexify certain semidefinite programs by converting them into an LMI.

\begin{prop}\label{prop:schur}
Let $A,B,C$ be matrices of compatible dimensions such that we can define the block matrix:
$X \defeq \left(\begin{smallmatrix} A & B \\ B^\tp & C \end{smallmatrix}\right)$
and suppose $X=X^\tp$ and $C \succ 0$. Then,
$
X \succeq 0$
 if and only if
$A-BC^{-1} B^\tp \succeq 0
$.
The result also holds when $\succeq$ is replaced by $\succ$.
\end{prop}

The second fundamental result is called the \textit{matrix elimination lemma}, by Gahinet and Apkarian~\cite{gahinet1994linear}. This result applies to matrix inequalities with affine dependence on a matrix of decision variables and is useful for eliminating controller parameters to obtain results independent of controller state dimension. 
\begin{lem}[\!\!\cite{gahinet1994linear}]\label{lem:matrix_elimination}
Given a symmetric matrix $\Psi \in \R^{n\times n}$ and two matrices $P$, $Q$ of column dimension $n$, consider the problem of finding some matrix $\Theta$ of compatible dimensions such that
\begin{equation}\label{eq:GA1}
\Psi + P^\tp \Theta^\tp Q  + Q^\tp \Theta P \prec 0.
\end{equation}
Denote by $W_P$, $W_Q$ any matrices whose columns form bases for the null spaces of $P$ and $Q$ respectively. Then there exists $\Theta$ satisfying~\eqref{eq:GA1} if and only if
\[
W_P^\tp \Psi W_P \prec 0 \quad\text{and}\quad W_Q^\tp \Psi W_Q \prec 0.
\] 
\end{lem}

The third fundamental result we will use is called the \textit{matrix completion lemma}, by Packard et al.~\cite[Lem.~7.5]{packard1991collection}. This result describes how a pair of matrices can be augmented to satisfy an inverse relationship.
\begin{lem}[\!\!\cite{packard1991collection}]\label{lem:matrix_completion}
Suppose $X,Y\in\R^{n\times n}$, with $X=X^\tp\succ 0$ and $Y=Y^\tp\succ 0$. Let $m$ be a positive integer. Then there exist matrices $X_2\in\R^{n\times m}$, $X_3\in\R^{m\times m}$ such that $X_3=X_3^\tp$, and
\[
\bmat{X & X_2 \\ X_2^\tp & X_3} \succ 0
\quad\text{and}\quad
\bmat{X & X_2 \\ X_2^\tp & X_3}^{-1} = \bmat{ Y & ? \\ ? & ? }
\]
if and only if $X-Y^{-1} \succeq 0$, and $\rank(X-Y^{-1}) \le m$.
\end{lem}

\begin{rem}
Applying~Proposition~\ref{prop:schur}, the condition in Lemma~\ref{lem:matrix_completion} can be expressed as an LMI in $X$ and $Y$:
\[
X - Y^{-1} \succeq 0
\quad\text{if and only if}\quad
\bmat{X & I \\ I & Y} \succeq 0.
\]
Also note that the condition $rank(X-Y^{-1})\le m$ is trivially satisfied for any $m\ge n$.
\end{rem}
  
\paragraph{Exponential stability with IQCs.} Finally, we will make use of an LMI for bounding the worst-case exponential convergence rate of a linear dynamical system subject to integral quadratic constraints (IQCs). Similar LMIs have appeared in~\cite{seiler2015stability,hu2016exponential,boczar2017exponential}.

\begin{lem}\label{lem:IQCLMI}
Consider the discrete-time LTI system $G$, which maps $u\mapsto y$ and obeys the dynamics:
\begin{subequations}\label{eq:lem_G}
\begin{align}
\xi_{k+1} &= A_G \xi_k + B_G u_k \\
y_k &= C_G \xi_k + D_G u_k
\end{align}
\end{subequations}
Fix $\rho \in (0,1]$, and consider the auxiliary system $\Psi_\rho$, which maps $(y,u)\mapsto z$ and obeys the dynamics:
\begin{subequations}\label{eq:lem_Psi}
\begin{align}
\zeta_{k+1} &= A_\Psi \zeta_k + B_{\Psi1} y_k + B_{\Psi2} u_k \\
        z_k &= C_\Psi \zeta_k + D_{\Psi1} y_k + D_{\Psi2} u_k
\end{align}
\end{subequations}
Suppose $(y,u)$ satisfies the $\rho$-\textit{Integral Quadratic Constraint} ($\rho$-IQC) defined by $(\Psi_\rho,M)$. That is,
\begin{equation}\label{eq:lem_IQC}
\sum_{k=0}^T \rho^{-2k} z_k^\tp M z_k \ge 0\quad\text{for }T=0,1,\dots
\end{equation}
Define state-space matrices for the map $u\mapsto z$:
\[
\stsp{A}{B}{C}{D}
\defeq \left[\begin{array}{cc|c}
A_G & 0 & B_G \\ B_{\Psi1} C_G & A_\Psi & B_{\Psi2}+B_{\Psi1}D_G \\ \hlinet
D_{\Psi1} C_G & C_\Psi & D_{\Psi2}+D_{\Psi1}D_G \\
\end{array}\right].
\]
If there exists some $P \succ 0$ such that
\begin{multline}\label{eq:lem_LMI}
\bmat{A & B}^\tp P \bmat{A & B} - \rho^2 \bmat{I & 0 }^\tp P \bmat{I & 0} \\
+ \bmat{C & D}^\tp M \bmat{C & D} \preceq 0,
\end{multline}
then the state of $G$ converges to zero at an exponential rate of $\rho$. Specifically, $\lim_{k\to \infty} \rho^{-k} \xi_k = 0$ for all initial states $(\xi_0,\zeta_0)$.
\end{lem}

\section{Algorithms as control systems}\label{sec:algo}

\begin{figure}[ht]
\centering
\includegraphics{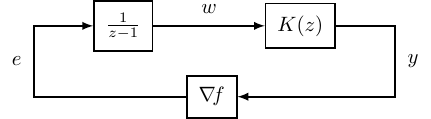}
\caption{Block diagram representing an iterative optimization algorithm. The integrator is present so that the algorithm can converge with no steady-state error for any initial condition of $K$.}\label{fig:block1}
\end{figure}

Following the line of research developed in~\cite{lessard16,cint_ACC}, we formulate the problem of iterative algorithm analysis as a control problem. Specifically, we will consider algorithms set up as in Figure~\ref{fig:block1}.

The iterative algorithm must contain a pure integrator, i.e. it must
take the form $K(z) \frac{1}{z-1}$ where $K(z)$ is an LTI system that
represents the algorithm.  Assume $K(z)$ has a state-space realization
$(A_K, B_K, C_K, D_K)$.  Let $w\in \R$ and $q \in \R^{n_K}$ be the
state of integrator and $K(z)$, respectively. The order of $K(z)$,
denoted $n_K$, is unspecified at this point, i.e. the algorithm may
have a finite but arbitrary amount of memory. A realization for
$K(z) \frac{1}{z-1}$ from input $e_k$ to output $y_k$ is given by:
\begin{subequations}\label{eq:KIntSS}
\begin{align}
  w_{k+1} &= w_k + e_k \\
  q_{k+1} &= A_K q_k + B_K w_k \\
  y_k &= C_K q_k + D_K w_k
\end{align}
\end{subequations}
This representation is very general. For example, the heavy-ball method is represented by $K(z) = \frac{-\alpha z}{(z-\beta)}$, which corresponds to $A_K=C_K=\beta$ and $B_K=D_K=-\alpha$. Meanwhile, the gradient descent algorithm corresponds to the static gain $K(z) = -\alpha$.

\paragraph{Dimensionality reduction.}
Throughout this paper, we assume for convenience that $K(z)$ is a single-input, single-output (SISO) system. For algorithms such as heavy-ball~\eqref{eq:heavyball} where $\alpha$ and $\beta$ are scalars, this assumption is not restrictive. Indeed, the transfer function for the heavy-ball algorithm takes the form $I_d \otimes K(z)\tfrac{1}{z-1}$, where $x_k \in \R^d$ in~\eqref{eq:heavyball}. The IQCs we will use to describe $\grad f$ and hence the LMIs~\eqref{eq:lem_LMI} will also have a block-Kronecker form. These facts allow us to restrict our attention to SISO $K(z)$ without any loss of generality. For further discussion on this type of dimensionality reduction, see~\cite[\S4.2]{lessard16}.

\begin{figure}[ht]
\centering
\includegraphics{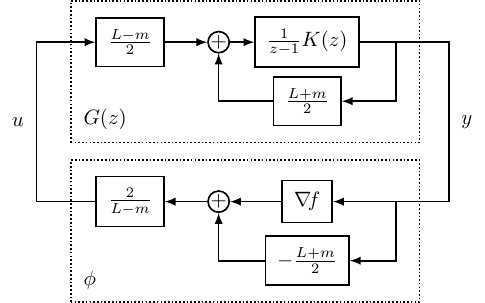}
\caption{Loop Transformation. If $\grad f$ is in the sector $(m,L)$, then $\phi$ is in the sector $(-1,1)$.}
\label{fig:SectorLoopTransf}
\end{figure}

\paragraph{Loop transformation.}
The case of interest in this paper is where the function $f$ is strongly convex and has Lipschitz gradients~\eqref{eq:convexity_assumptions}. This implies, in the language of robust control, that $\grad f$ is sector-bounded and slope-restricted in the interval $(m,L)$. In order to streamline the rest of the paper, we will perform a loop transformation to shift this interval to $(-1,1)$.
The transformation is shown in Figure~\ref{fig:SectorLoopTransf}. 

The loop-shifting transformation normalizes the nonlinearity $\phi$ to lie in the
sector $(-1,1)$. As a result, $u_k=\phi(y_k)$
satisfies the pointwise-in-time constraint:
\begin{equation}\label{sector_bound}
\bmat{ y_k \\ u_k}^\tp \bmat{1 & 0 \\ 0 & -1} \bmat{y_k \\ u_k} \ge 0.
\end{equation}
The input to $K(z) \frac{1}{z-1}$ is transformed by
the following algebraic equation:
$ e_k = \tfrac{L-m}{2} u_k + \tfrac{L+m}{2} y_k $.
A realization for the transformed system $G(z)$ is obtained by
combining this relation with \eqref{eq:KIntSS}.  Eliminating $e_k$ 
yields the following state-space realization with state $(q_k,w_k)$:
\begin{equation}\label{eq:G}
G(z) \defeq \squeeze{1pt}\stsp{A_G}{B_G}{C_G}{D_G}
=
\squeeze{1pt}
\left[\begin{array}{cc|c}
A_K & B_K & 0 \\
\tfrac{L+m}{2} C_K & 1\!+\!\tfrac{L+m}{2} D_K & \tfrac{L-m}{2} \\ \hline
C_K & D_K & 0
\end{array}\right]
\end{equation}


\section{Sector-bounded nonlinearities}\label{sec:sector}

In this section we treat the analysis problem for the special case where $\grad f$ is \textit{sector-bounded}. This assumption corresponds to relaxing~\eqref{eq:convexity_assumptions} to hold only between an arbitrary point and the optimizer of $f$. In other words,~\eqref{eq:convexity_assumptions} reduces to:
\begin{align*}\label{eq:sector_assumptions}
\bmat{\grad f(x) \\ x-x^\star}^{\!\tp}\!\!
\squeeze{3pt}\bmat{ -2mL & L+m \\ L+m & -2 }\!\!
\bmat{\grad f(x) \\ x-x^\star} \ge 0
\quad \text{for all }x\in\R^n.
\end{align*}
From the IQC standpoint, this means that we assume the transformed nonlinearity $\phi$ only satisfies the pointwise-in-time constraint~\eqref{sector_bound}.

Our goal is to find an algorithm $(A_K,B_K,C_K,D_K)$ such
that the convergence rate $\rho$ is minimized. The sector IQC~\eqref{sector_bound} has no dynamics, so let $z = \left(\begin{smallmatrix}y \\ u\end{smallmatrix}\right)$. In this case, $\Psi=I_2$ and the state matrices for the map
$u\mapsto z$ in Lemma~\ref{lem:IQCLMI} simplify to:
\[
\stsp{A}{B}{C}{D}
\defeq 
\stsp{A_G}{B_G}{\bsmat{C_G \\ 0}}{\bsmat{D_G \\ 1}}.
\]
By Lemma~\ref{lem:IQCLMI}, iterates
converge with rate $\rho\in(0,1]$ if there exists $P\succ 0$ such that:
\begin{multline}
\bmat{A_G & B_G}^\tp P \bmat{A_G & B_G} - \rho^2 \bmat{I & 0 }^\tp P \bmat{I & 0} \\
+ \bmat{C_G & D_G \\ 0 & 1}^\tp \bmat{1 & 0 \\ 0 & -1} \bmat{C_G & D_G \\ 0 & 1} \preceq 0.
\end{multline}
We can rearrange the LMI above to the form:
\begin{equation}\label{eq:LMI2}
\bmat{A_G & B_G \\ C_G & D_G}^\tp \bmat{ P & 0 \\ 0 & 1} 
\bmat{A_G & B_G \\ C_G & D_G}
- \bmat{\rho^2 P & 0 \\ 0 & 1} \preceq 0.
\end{equation}
The design corresponds to synthesizing the algorithm 
$(A_K,B_K,C_K,D_K)$, Lyapunov matrix $P\succ 0$, and convergence
rate bound $\rho\in(0,1]$.  The matrix inequality~\eqref{eq:LMI2}
is nonlinear due to products of these variables. Applying Proposition~\ref{prop:schur}, we can write the equivalent condition:
\begin{equation}\label{eq:schur1}
\bmat{ \rho^2 P & 0 & A_G^\tp & C_G^\tp \\ 0 & 1 & B_G^\tp & D_G^\tp \\
A_G & B_G & P^{-1} & 0 \\ C_G & D_G & 0 & 1} \succeq 0,
\quad\text{and}\quad P \succ 0.
\end{equation}
From~\eqref{eq:G}, we can write $(A_G,B_G,C_G,D_G)$ as an affine function of the controller parameters. In the equation below, dashed lines indicate block structure (not state-space notation).
\begin{multline}\label{eq:CLsub}
\left[\begin{array}{c;{2pt/2pt}c}A_G & B_G \\ \hdashline[2pt/2pt] C_G & D_G\end{array}\right]
= \left[\begin{array}{cc;{2pt/2pt}c}0 & 0 & 0 \\ 0 & 1 & \tfrac{L-m}{2} \\ \hdashline[2pt/2pt]
0 & 0 & 0
\end{array}\right] \\
+
\left[\begin{array}{cc}I & 0 \\ 0 & \tfrac{L+m}{2} \\ \hdashline[2pt/2pt] 0 & 1 \end{array}\right]
\underbrace{\bmat{A_K & B_K \\ C_K & D_K}}_{K}
\left[\begin{array}{cc;{2pt/2pt}c}I & 0 & 0 \\ 0 & 1 & 0
\end{array}\right]
\end{multline}
Substituting~\eqref{eq:CLsub} into~\eqref{eq:schur1}, we obtain:
\begin{multline}\label{eq:schurexp}
\underbrace{\bmat{ \rho^2 P & \bsmat{0\\0} & \bsmat{0 & 0 \\ 0 & 1} & \bsmat{0 \\ 0} \\[1mm]
\bsmat{0 & 0} & 1 & \bsmat{0 & \tfrac{L-m}{2}} & 0 \\[1mm]
\bsmat{0 & 0 \\ 0 & 1} & \bsmat{0 \\ \tfrac{L-m}{2}} & P^{-1} & \bsmat{0\\0} \\
\bsmat{0 & 0} & 0 & \bsmat{0 & 0} & 1 }}_{\Psi}
\\
+
\sym \bmat{ 0 & 0 \\ 0 & 0 \\ \bsmat{I \\ 0} & \bsmat{ 0 \\ \tfrac{L+m}{2} } \\ 0 & 1 }K
\bmat{ \bsmat{I & 0 } & 0 & 0 & 0 \\[1mm] \bsmat{ 0 & 1 } & 0 & 0 & 0 } \succeq 0,
\end{multline}
where $\sym$ is the shorthand notation we defined in Section~\ref{sec:prelim} and $K$ is defined in~\eqref{eq:CLsub}.
By Lemma~\ref{lem:matrix_elimination}, the LMI~\eqref{eq:schurexp} is feasible if and only if a pair of conditions hold. In this case, the conditions are:
\begin{align*}
\squeeze{2pt}
\bmat{ I & 0 & \bsmat{0 & 0} & 0 \\[1mm] 0 & 1 & \bsmat{0 & 0} & 0 \\[1mm]
0 & 0 & \bsmat{0 & 1} & -\tfrac{L+m}{2} }
\Psi
\bmat{I & 0 & \bsmat{0 & 0} & 0 \\[1mm] 0 & 1 & \bsmat{0 & 0} & 0 \\[1mm]
0 & 0 & \bsmat{0 & 1} & -\tfrac{L+m}{2} }^\tp
&\succeq 0 \\
\bmat{ \bsmat{0\\0} & \bsmat{0\\0} & \bsmat{0\\0} \\
1 & 0 & 0 \\ 0 & I & 0 \\ 0 & 0 & 1 }^\tp
\Psi
\bmat{ \bsmat{0\\0} & \bsmat{0\\0} & \bsmat{0\\0} \\
1 & 0 & 0 \\ 0 & I & 0 \\ 0 & 0 & 1 }
&\succeq 0
\end{align*}
where $\Psi$ is defined in~\eqref{eq:schurexp}.
Carrying out the matrix multiplications and simplifying yields:
\begin{subequations}\label{eq:GahApk}
\begin{align}
\label{eq:GahApk1}
\bmat{ \rho^2 P & \bsmat{0\\0} & \bsmat{0\\1} \\[1mm]
\bsmat{ 0 & 0 } & 1 & \tfrac{L-m}{2} \\[1mm]
\bsmat{0 & 1 } & \tfrac{L-m}{2}  & \bsmat{0\\1}^\tp\! P^{-1}\! \bsmat{0\\1} +(\tfrac{L+m}{2})^2 } & \succeq 0,\\
\label{eq:GahApk2}
\bmat{ 1 & \bsmat{0 & \tfrac{L-m}{2}} & 0 \\ \bsmat{ 0 \\ \tfrac{L-m}{2}} & P^{-1} & \bsmat{0\\0} \\ 0 & \bsmat{0 & 0} & 1} &\succeq 0
\end{align}
\end{subequations}
Applying Proposition~\ref{prop:schur} to the first $2\times 2$ block of~\eqref{eq:GahApk1} we obtain the following equivalent condition:
\begin{align*}
r + (\tfrac{L+m}{2})^2
- \left( \rho^{-2} r + (\tfrac{L-m}{2})^2 \right) &\ge 0 
\end{align*}
where $r \defeq  \bsmat{0\\1}^\tp\! P^{-1}\! \bsmat{0\\1} > 0$. This
simplifies to:
\begin{equation}
\label{eq:GahApk1Simp}
mL \ge (\rho^{-2}-1) r
\end{equation}
Next, Equation~\eqref{eq:GahApk2} is feasible if and only if the first
$2\times 2$ block is feasible.  Applying Proposition~\ref{prop:schur} to
the upper left corner of this $2 \times 2$ block yields:
$r \ge (\tfrac{L-m}{2})^2$.
Combining with~\eqref{eq:GahApk1Simp} 
yields $4mL \ge (\rho^{-2} - 1)(L-m)^2$. Solving for $\rho$ yields:
$\rho \ge \frac{L-m}{L+m}$, which is precisely the optimal gradient
rate bound described in Section~\ref{sec:intro}. Indeed, if we also
set $P^{-1} = \bmat{ \epsilon & 0 \\ 0 & (\frac{L-m}{2})^2 }$ for any
$\epsilon > 0$, Equation~\eqref{eq:GahApk} is satisfied. Setting
$D_K = -\frac{2}{m+L}$ (and $A_K$ empty), which is the gradient method
with optimal stepsize, renders~\eqref{eq:schurexp} feasible too.

\paragraph{Discussion.}

It is known that for gradient descent, choosing $\alpha=\tfrac{2}{m+L}$ optimizes the rate bound~\eqref{eq:gradrate} for strongly convex functions with Lipschitz continuous gradients. Similar results have shown that the same upper bound (and optimal stepsize) holds if we constrain $f$ to be quadratic or if we relax $f$ to merely have sector-bounded gradients~\cite{lessard16}. It is also straightforward to show that the rate bound of $\tfrac{L-m}{L+m}$ is tight, because it is attained when using this algorithm to optimize the scalar quadratic function $f(x) \defeq \tfrac{m}{2} (x-x^\star)^2$.

This bound assumes the algorithm in question is gradient descent. In this section, we showed that for the class of functions with sector-bounded gradients, gradient descent with $\alpha=\tfrac{2}{L+m}$ is in fact optimal over all algorithms that can be expressed as LTI systems (of any fixed order) in feedback with $\grad f$, as in Figure~\ref{fig:SectorLoopTransf}. Put another way, when using sector-bounded gradients, we cannot improve our upper bound by using more complex algorithms.


\section{Sector-bounded and slope restricted}\label{sec:sloperestricted}

We now extend our approach from Section~\ref{sec:sector} to the case where $\grad f$ is both sector-bounded and \textit{slope-restricted}. Specifically, we assume that $f$ is strongly convex with Lipschitz-continuous gradients so that $\grad f$ satisfies the constraints
in~\eqref{eq:convexity_assumptions}.  The transformed nonlinearity $\phi$ thus satisfies the normalized constraint for all $x,y\in\R^n$:
\begin{align*}
  \bmat{x-y \\ \phi(x)-\phi(y)}^{\tp}
  \bmat{ 1 & 0 \\ 0 & -1 }
  \bmat{x-y \\ \phi(x)-\phi(y)} \ge 0.
\end{align*}
Such functions satisfy a general class of $\rho$-IQCs defined by
Zames--Falb multipliers
\cite{carrasco2016zames,boczar2017exponential,freeman2018noncausal}. For now, we will assume a general IQC of the form~\eqref{eq:lem_Psi}, and we further break apart $z$ into its two components, thereby resulting in joint plant-IQC dynamics:
\begin{align*}
\bmat{\xi_{k+1} \\ \zeta_{k+1}}
&= \bmat{A_G & 0 \\ B_\Psi^1 C_G & A_\Psi}\bmat{\xi_k \\ \zeta_k} 
+ \bmat{ B_G \\ B_\Psi^1 D_G + B_\Psi^2} u_k \\
\bmat{z_k^1\\z_k^2} &= \bmat{ D_\Psi^{11} C_G & C_\Psi^1 \\ D_\Psi^{21} C_G & C_\Psi^2 }\bmat{\xi_k \\ \zeta_k} + \bmat{ D_\Psi^{11} D_G + D_\Psi^{12}\\ D_\Psi^{21} D_G + D_\Psi^{22}} u_k
\end{align*}
Substituting $(A_G,B_G,C_G,D_G)$ from~\eqref{eq:G}, obtain:
\begin{subequations}\label{eq:CLpsi}
\begin{align}
\label{eq:CLpsia}
\bmat{q_{k+1} \\ w_{k+1} \\ \zeta_{k+1}} &\!=\!
\squeeze{3pt}
\bmat{ A_K & B_K & 0 \\ \tfrac{L+m}{2} C_K & 1\!+\!\tfrac{L+m}{2} D_K & 0 \\ B_\Psi^1 C_K & B_\Psi^1 D_K & A_\Psi}\!
\bmat{q_k \\ w_k \\ \zeta_k}
\!+\!
\bmat{0 \\ \tfrac{L-m}{2} \\ B_\Psi^2} \!u_k\\
\label{eq:CLpsib}
\bmat{z_k^1\\z_k^2} &\!=\! \bmat{ D_\Psi^{11} C_K & D_\Psi^{11} D_K & C_\Psi^1 \\ D_\Psi^{21} C_K & D_\Psi^{21} D_K & C_\Psi^2 }\!\bmat{q_k \\ w_k \\ \zeta_k}
\!+\! \bmat{ D_\Psi^{12}\\ D_\Psi^{22}} \!u_k
\end{align}
\end{subequations}
This closed loop includes the algorithm (integrator and controller) and the IQC states.  The closed-loop state matrices are denoted $\left( A,B,\bmat{C_1\\ C_2},\bmat{D_1\\D_2}\right)$.  In this case the LMI in Lemma~\ref{lem:IQCLMI} looks different because the IQC term is quadratic in $(z^1,z^2)$ instead of $(y,u)$. We obtain:
\begin{multline*} 
\bmat{A & B \\ C_1 & D_1}^\tp \bmat{ P & 0 \\ 0 & 1} \bmat{A & B \\ C_1 & D_1}\\
- \bmat{I & 0 \\ C_2 & D_2}^\tp\bmat{\rho^2 P & 0 \\ 0 & 1}\bmat{I & 0 \\ C_2 & D_2} \preceq 0
\end{multline*}
As before, we can use Proposition~\ref{prop:schur} to convert this to a single LMI:
\begin{equation}\label{eq:schur2}
\bmat{ \rho^2 P +C_2^\tp C_2 & C_2^\tp D_2 & A^\tp & C_1^\tp \\
D_2^\tp C_2 & D_2^\tp D_2 & B^\tp & D_1^\tp \\
A & B & P^{-1} & 0 \\ C_1 & D_1 & 0 & 1} \succeq 0,\quad P \succ 0
\end{equation}
Equation~\eqref{eq:schur2} will be affine in the algorithm parameters $(A_K,B_K,C_K,D_K)$ and everything will carry through as before if $C_2$ and $D_2$ do not depend on the algorithm parameters. If we examine the formulas of~\eqref{eq:CLpsib}, this amounts to requiring for example that $D_\Psi^{21}=0$ and $D_\Psi^{22}=1$. So the question becomes: is it possible to find a factorization $\Psi$ of the IQC with this property? 

The general Zames--Falb $\rho$-IQC for a nonlinearity that is sector-bounded and slope-restricted on $(-1,1)$ is:
\[
\Pi(z) \defeq 
\frac{1}{2}\bmat{2-\hat h -\hat h^* & \hat h - \hat h^* \\
\hat h^* - \hat h & -2+\hat h+\hat h^*}
\]
Where $\hat h(z)$ is a discrete-time transfer function with impulse response $h_k$ satisfying:
\[
\sum_{k=-\infty}^\infty \max\bigl(1,\rho^{-2k}\bigr)|h_k| \le 1
\]
If we assume for now that the Zames--Falb $\rho$-IQC is causal, then a possible factorization is:
\begin{gather*}
\Pi(z) = \Psi(z)^*\! \bmat{1 & 0 \\ 0 & -1} \!\Psi(z),\text{ with }
\Psi(z) = \frac{1}{2}\!\squeeze{3pt}\bmat{2\!-\!\hat h & \hat h \\ \hat h & 2\!-\!\hat h }\!.
\end{gather*}
Assume $\hat{h}$ is a finite impulse response filter, i.e. $\hat h(z) = h_1 z^{-1} + h_2 z^{-2} + \cdots + h_k z^{-k}$. Consider the realization:
\[
\Psi(z) = \left[\begin{array}{cccc|cc}
0 & 1 & \cdots & 0 & 0 & 0 \\
0 & 0 & \ddots & \vdots & \vdots & \vdots \\
\vdots & \vdots & \ddots & 1 & 0 & 0 \\
0 & 0 & \cdots & 0 & \tfrac{1}{2} & -\tfrac{1}{2} \\ \hline
-h_k & -h_{k-1} & \cdots & -h_1 & 1 & 0 \\
h_k & h_{k-1} & \cdots & h_1 & 0 & 1 
\end{array}\right]
\]
We see that $D_\Psi^{21}=0$ and $D_\Psi^{22}=1$, as desired. Therefore, Equation~\eqref{eq:schur2} holds, but the $\hat h$ coefficients only appear in $C_1$ and $C_2$, and the algorithm parameters only appear in an affine fashion and in the terms $A, B, C_1, D_1$. The lower-left corner of~\eqref{eq:schur2} can be written in terms of the algorithm parameters by substituting~\eqref{eq:CLpsi}. The result is (again, dashed lines indicate block structure)
\begin{multline}\label{eq:decomp} 
\left[\begin{array}{c;{2pt/2pt}c}A & B \\ \hdashline[2pt/2pt] C_1 & D_1 \end{array}\right]
= \underbrace{\left[\begin{array}{ccc;{2pt/2pt}c}
0 & 0 & 0 & 0 \\
0 & 1 & 0 & \tfrac{L-m}{2} \\
0 & 0 & A_\Psi & B_\Psi^2 \\ \hdashline[2pt/2pt]
0 & 0 & C_\Psi^1 & D_\Psi^{12} 
\end{array}\right]}_\Phi \\
 + \underbrace{\left[\begin{array}{cc}
I & 0 \\ 0 & \tfrac{L+m}{2} \\ 0 & B_\Psi^1 \\ \hdashline[2pt/2pt] 0 & D_\Psi^{11}
\end{array}\right] }_S
\underbrace{\bmat{A_K & B_K \\ C_K & D_K}}_K
\underbrace{\left[\begin{array}{ccc;{2pt/2pt}c}
I & 0 & 0 & 0 \\ 0 & 1 & 0 & 0 \end{array}\right]}_T
\end{multline}
Let columns of $S^\perp$ and $T^\perp$ be bases for the left and right nullspaces of $S$ and $T$, respectively.  These are given by:
\begin{align*}
S^\perp &= \bmat{I & 0 \\ 0 & \tfrac{L+m}{2} \\ 0 & B_\Psi^1 \\ 0 & 1 }^\perp
= \bmat{0 & 1 & 0 & -\frac{L+m}{2} \\ 0 & 0 & B^\perp_1 & B^\perp_2}, \\
T^\perp &= \bmat{ I & 0 & 0 & 0 \\ 0 & 1 & 0 & 0}^\perp = \bmat{0 & 0 \\ 0 & 0 \\ I & 0 \\ 0 & 1}
\end{align*}
where $\bmat{B_1^\perp & B_2^\perp}\bmat{B_\Psi^1 \\ 1} = 0$. Substituting~\eqref{eq:decomp} into~\eqref{eq:schur2} and applying Lemma~\ref{lem:matrix_elimination}, the LMI condition~\eqref{eq:schur2}
is feasible if and only if:
\begin{align*}
\squeeze{3pt}
\bmat{ I & 0 \\ 0 & S^\perp}\bmat{ \bmat{\rho^2 P +C_2^\tp C_2 & C_2^\tp \\ C_2 & 1}  & \Phi^\tp \\
\Phi &  \bmat{ P^{-1} & 0 \\ 0 & 1} }\bmat{ I & 0 \\ 0 & S^\perp}^\tp &\succeq 0 \\
\squeeze{3pt}
\bmat{ T^\perp & 0 \\ 0 & I}^\tp\bmat{ \bmat{\rho^2 P +C_2^\tp C_2 & C_2^\tp \\ C_2 & 1}  & \Phi^\tp \\
\Phi &  \bmat{ P^{-1} & 0 \\ 0 & 1} }\bmat{ T^\perp & 0 \\ 0 & I} &\succeq 0
\end{align*}
Expanding and factoring the (1,1) block to expose $P$:
\begin{align*}
\squeeze{2pt}\bmat{ \bmat{I & 0 \\ C_2 & 1}^\tp\bmat{\rho^2 P & 0 \\ 0 & 1}\bmat{I & 0 \\ C_2 & 1} & \Phi^\tp S^{\perp\tp} \\ S^\perp \Phi & S^\perp\bmat{ P^{-1} & 0 \\ 0 & 1}S^{\perp\tp} } &\succeq 0\\
\squeeze{2pt}\bmat{ T^{\perp\tp}\bmat{I & 0 \\ C_2 & 1}^\tp\bmat{\rho^2 P & 0 \\ 0 & 1}\bmat{I & 0 \\ C_2 & 1}T^\perp & T^{\perp\tp}\Phi^\tp \\ \Phi T^\perp & \bmat{ P^{-1} & 0 \\ 0 & 1} } &\succeq 0
\end{align*}
Apply Proposition~\ref{prop:schur} to the (1,1) block of the first LMI above
to obtain an equivalent LMI in $Q \defeq P^{-1}$. Next
apply Proposition~\ref{prop:schur} to the (2,2) block of the second LMI above
to obtain an equivalent LMI in $P$.  This yields the following two conditions:
\begin{multline*}
S^\perp \bmat{ Q & 0 \\ 0 & 1}S^{\perp\tp} \\
-\squeeze{2pt} S^\perp \Phi  \bmat{I & 0 \\ -C_2 & 1}\bmat{\rho^{-2} Q & 0 \\ 0 & 1}\bmat{I & 0 \\ -C_2 & 1}^\tp\Phi^\tp S^{\perp\tp} \succeq 0
\end{multline*}
\vspace{-1em}
\begin{multline*}
T^{\perp\tp}\bmat{I & 0 \\ C_2 & 1}^\tp\bmat{\rho^2 P & 0 \\ 0 & 1}\bmat{I & 0 \\ C_2 & 1}T^\perp\\
- T^{\perp\tp}\Phi^\tp \bmat{ P & 0 \\ 0 & 1} \Phi T^\perp \succeq 0
\end{multline*}
These LMIs are block $4\times 4$ and the dimensions are $(n_K + 1 + n_\Psi + 1)$.
To proceed further, we need to break $P$ and $Q$ into their $3\times 3$ blocks. These blocks have dimensions $(n_K + 1 + n_\Psi)$. Substitute for
$(S^\perp,T^\perp)$ and note that $C_2$ simplifies to:
$
C_2 = \bmat{ D_\Psi^{21} C_K & D_\Psi^{21} D_K & C_\Psi^2 }
= \bmat{ 0 & 0 & C_\Psi^2 } 
$. Upon further simplification, we find the first block-rows and block-columns of $P$ and $Q$ (which had dimension $n_K$ cancel out. The resulting LMIs are:
\begin{subequations}
\label{eq:final}
\begin{multline}
 \bmat{1 & 0 & -\frac{L+m}{2} \\ 0 & B^\perp_1 & B^\perp_2} \bmat{ Q_{22} & Q_{23} & 0 \\ Q_{32} & Q_{33} & 0 \\ 0 & 0 & 1} \bmat{\star}^\tp \\
 -  \bmat{1 & 0 & -\frac{L+m}{2} \\ 0 & B^\perp_1 & B^\perp_2}
 \bmat{1 & 0 & \tfrac{L-m}{2} \\ 0 & A_\Psi & B_\Psi^2 \\ 0 & C_\Psi^1 & D_\Psi^{12}}
 \bmat{1 & 0 & 0 \\ 0 & I & 0 \\ 0 & -C_\Psi^2 & 1} \\
 \times
 \bmat{ \rho^{-2}Q_{22} & \rho^{-2}Q_{23} & 0 \\ \rho^{-2}Q_{32} & \rho^{-2}Q_{33} & 0 \\ 0 & 0 & 1}\bmat{\star}^\tp \succeq 0
\end{multline}
\vspace{-1em}
\begin{multline}
\bmat{\star}^\tp \bmat{ \rho^{2}P_{22} & \rho^{2}P_{23} & 0 \\ \rho^{2}P_{32} & \rho^{2}P_{33} & 0 \\ 0 & 0 & 1}
\bmat{1 & 0 & 0 \\ 0 & I & 0 \\ 0 & C_\Psi^2 & 1}
\bmat{0 & 0 \\ I & 0 \\ 0 & 1} \\
-\bmat{\star}^\tp \bmat{ P_{22} & P_{23} & 0 \\ P_{32} & P_{33} & 0 \\ 0 & 0 & 1}
\bmat{1 & 0 & \tfrac{L-m}{2} \\ 0 & A_\Psi & B_\Psi^2 \\ 0 & C_\Psi^1 & D_\Psi^{12}}
\bmat{0 & 0 \\ I & 0 \\ 0 & 1} \succeq 0
\end{multline}
These conditions are two LMIs involving (sub-blocks) of $P$ and $Q$ but
coupled by the non-convex constraint $PQ=I$. Apply
Lemma~\ref{lem:matrix_completion} to replace the non-convex constraint
by a rank constraint and the following LMI:
\begin{equation}
\bmat{ P_{22} & P_{23} & 1 & 0 \\
       P_{32} & P_{33} & 0 & I \\
       1 & 0 & Q_{22} & Q_{23} \\
       0 & I & Q_{32} & Q_{33} } \succeq 0
\end{equation}
\end{subequations}
The rank constraint is trivially satisfied if $n_K\ge 2$ and hence can
be neglected. In this case \eqref{eq:final} provides a set of three
LMI conditions in $P$ and $Q$. In particular, $n_K=2$ corresponds to a
third-order algorithm (two states in $K$ plus the additional
integrator).

\paragraph{Discussion.}
For each fixed value of $\rho$, if~\eqref{eq:final} is feasible for some $P$ and $Q$, then it is possible to synthesize a controller $(A_K,B_K,C_K,D_K)$ for which the rate of convergence $\rho$ is certified over the class of sector-bounded and slope-restricted functions.

Unlike previous analyses where the algorithm is fixed and one seeks to certify a certain convergence rate by searching over IQCs of a given class~\cite{ralg_ACC,lessard16,diss_ICML,iqcadmm_ICML,distrop_Allerton}, here we fix the particular Zames-Falb IQC and instead certify a rate of convergence over all possible algorithms. This synthesis is performed numerically in the next section for the
class of off-by-one IQCs.

\section{Off-by-one IQC certification}

A special instance of the general class of Zames--Falb multipliers discussed in Section~\ref{sec:sloperestricted} is the weighted off-by-one
$\rho$-IQC \cite{lessard16}. For a given $\rho \in (0,1]$ and any
$h_1 \in [0,\rho^2]$, the function $\phi$ satisfies the $\rho$-IQC
defined by:
\begin{equation}\label{eq:offby1}
\begin{gathered}
   A_\Psi = 0, \quad
   B_\Psi = \bmat{ \tfrac{1}{2} & -\tfrac{1}{2} }, \quad
   C_\Psi = \bmat{ -h_1 \\  h_1}, \\
   D_\Psi = \bmat{ 1 & 0 \\ 0 &1 }, \quad
   M = \bmat{ 1 & 0 \\ 0 & -1 }.
\end{gathered}
\end{equation}
The filter $\Psi$ is parameterized by $\rho$ and the free variable
$h_1$.  It has one timestep of memory and the corresponding
$\rho$-IQC captures the slope conditions that must hold between data
$(y_k,u_k)=(y_k,\phi(y_k))$ at one timestep and the previous timestep
$(y_{k-1},u_{k-1})$. If $h_1=0$ then the constraint reduces to the
sector IQC. In other words, this form of the weighted off-by-1 subsumes the
sector bound used in Section~\ref{sec:sector}.  The more general Zames--Falb multipliers uses a filter $\Psi$ with additional memory to incorporate the slope constraints across multiple time steps.

\begin{figure}[htb]
\centering
\includegraphics{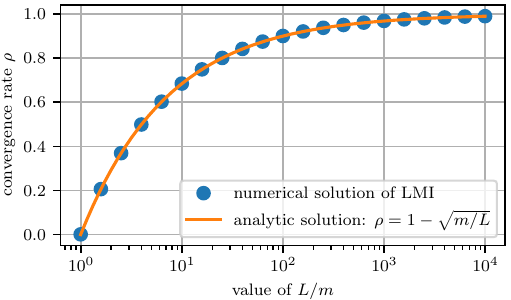}
\caption{The circular dots are numerical solutions of the LMI~\eqref{eq:final} for different values of $L/m$, using the off-by-one IQC and performing a bisection search on $\rho$. The solid curve is a plot of the function $\rho = 1-\sqrt{m/L}$, which is the fastest known achievable worst-case rate~\cite{van2018fastest}.}
\label{fig:numeric}
\end{figure}

Substituting the IQC~\eqref{eq:offby1} into the semidefinite program~\eqref{eq:final} and using $h_1 = \rho^2$ as in~\cite{lessard16}, we obtain a convex program in $(P,Q)$ for each fixed value of $\rho$ (and for fixed values of $L$ and $m$). Performing a bisection search on $\rho$, we can obtain the minimum $\rho$ such that the LMI is feasible. We plot this result in Figure~\ref{fig:numeric}, where we can see that the numerical results agree perfectly with $\rho = 1-\sqrt{m/L}$, which is the fastest known convergence rate achievable for strongly convex functions with Lipschitz-continuous gradients, and it is achieved by the \textit{triple momentum method}~\cite{van2018fastest}.

\section{Concluding remarks}\label{sec:conclusion}

We demonstrated that tools from robust synthesis can be leveraged for algorithm design. Unlike prior works that fix the algorithm size (fixed memory) and seek a Lyapunov candidate that guarantees a certain convergence rate over a class of functions, we instead fix the IQC used to characterize the class of functions, and then search over all possible algorithms (with finite but arbitrary memory) for the fastest certifiable rate.

We show that if $\grad f$ is sector-bounded, no algorithm of any size can outperform gradient descent. We then show that if $\grad f$ satisfies the weighted off-by-one $\rho$-IQC~\cite{lessard16}, no algorithm of any (finite) memory can outperform the triple momentum method.
In short, we show that eliminating the control variables from the associated LMIs can lead to novel algorithm performance bounds that are independent of algorithm dimension.


\begin{small}
\bibliographystyle{abbrv}
\bibliography{algo_synth}
\end{small}

\end{document}